\documentclass[journal=jctcce,manuscript=article]{achemso}

\usepackage[T1]{fontenc} 
\usepackage{tgpagella}   
\usepackage{microtype}   
\usepackage{hyperref}    
\usepackage{fontawesome} 

\usepackage{amsmath}    
\usepackage{cancel}     
\usepackage{mathtools}  
\usepackage{bm}         
\usepackage{physics}    
\usepackage{tensor}     
\usepackage{braket}     
\usepackage{booktabs}   

\usepackage[version=3]{mhchem} 

\usepackage{graphicx} 
\usepackage{dcolumn}  
\usepackage{multicol} 
\usepackage{subcaption}
\usepackage{float}

\usepackage{verbatim} 
\usepackage[x11names]{xcolor} 
\usepackage{soul}     

\usepackage{bbold}    
\usepackage{systeme}  
\usepackage{multirow}

\allowdisplaybreaks[1]
\graphicspath{{}}

\SectionNumbersOn

\author{Alberto Barlini}
\affiliation{Scuola Normale Superiore, Pisa, Italy}

\author{Andrea Bianchi}
\affiliation{Scuola Normale Superiore, Pisa, Italy}

\author{Jan Haakon Melka-Trabski}
\affiliation{Department of Chemistry, Norwegian University of Science and Technology, Trondheim, Norway}

\author{Julien Bloino}
\affiliation{Scuola Normale Superiore, Pisa, Italy}

\author{Henrik Koch}
\email{henrik.koch@ntnu.no}
\affiliation{Department of Chemistry, Norwegian University of Science and Technology, Trondheim, Norway}

\title[]{Cavity Field-Driven Symmetry Breaking and Modulation of Vibrational Properties: Insights from the Analytical QED-HF Hessian}

\abbreviations{ab initio QED, polaritonic chemistry, QED response theory, strong coupling, molecular properties}


\begin{document}


\begin{abstract}
    In this work, we present the analytical derivation and implementation of the quantum electrodynamics Hartree–Fock Hessian. We investigate how electronic strong coupling influences molecular vibrational properties, applying this framework to formaldehyde, p-nitroaniline, and adamantane. Our analysis reveals cavity-induced changes in vibrational frequencies and intensities. Additionally, we show how the quantum electromagnetic field breaks molecular symmetry, activating previously forbidden infrared transitions. Our findings highlight the potential of strong coupling as a method for controlling and modulating molecular vibrational properties.
\end{abstract}

\begin{figure}[H]
    \centering
    \includegraphics[width=0.55\linewidth]{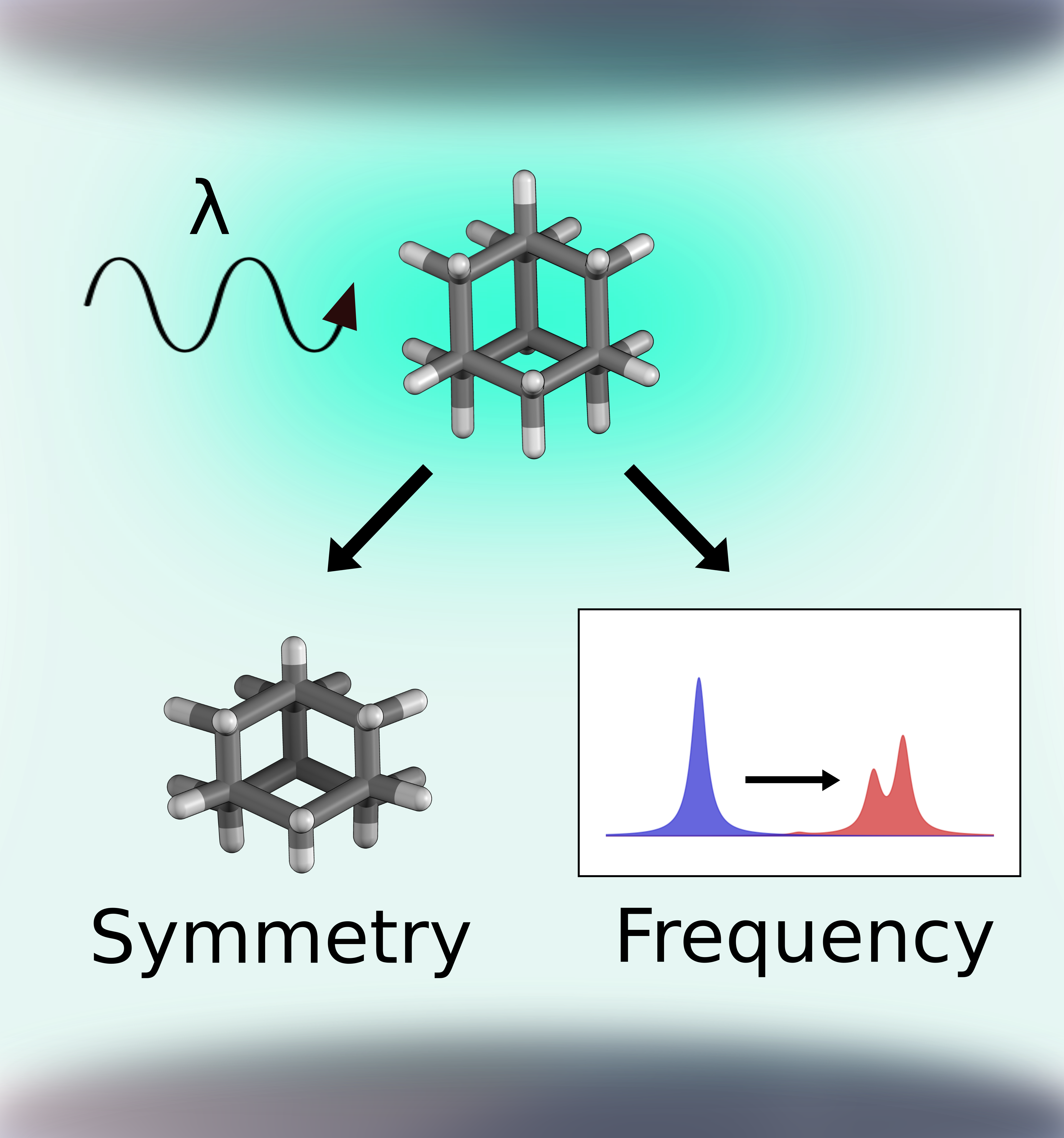}
\end{figure}

\section{Introduction}

Computing the analytical Hessian is essential for an accurate and numerically stable description of the vibrational properties and for identifying stationary points on the potential energy surface, enabling the detailed analysis of vibrational spectra and providing deeper insight into chemical reaction pathways. Significant effort has been directed towards developing theoretical methods and algorithms to evaluate the Hessian using various approximate wave function methods, which rely on the analytical second derivatives of the electronic energy \cite{pulay1969,bishop1966,stanton2000,handy1985,harrison1986,jorgensen89_mp2,helgaker1989,koch1990,kallay2004,gauss1997,gauss2000}.

In the strong light-matter coupling regime, interactions between molecules and confined electromagnetic fields form hybrid light-matter states known as polaritons. Theoretical descriptions of these effects are based on quantum electrodynamics (QED) extensions of electronic structure methods, including QED density functional theory (QEDFT/QED-DFT) \cite{ruggenthaler2014quantum, pellegrini2015}, QED Hartree–Fock (QED-HF) \cite{Tor2020,deprince2021cavity}, strong coupling QED Hartree-Fock (SC-QED-HF) \cite{riso2022molecular, el2024toward}, second-order QED M{\o}ller-Plesset perturbation theory (QED-MP2 and SC-QED-MP2) \cite{bauer2023perturbation, moutaoukal2025strong}, QED complete active space configuration interaction (QED-CASCI) \cite{vu2024}, QED density matrix renormalization group method (QED-DMRG) \cite{matusek2024}, and QED coupled cluster theory (QED-CC) \cite{Tor2020, mordovina2020polaritonic, pavosevic2021polaritonic}. These methods provide a framework for studying cavity-induced modifications to ground and excited-state molecular properties.\cite{castagnola2025strong} Recent studies have explored cavity-induced effects on molecular geometries and infrared (IR) spectra under vibrational strong coupling (VSC) \cite{schnappinger2023_ab_initio, schnappinger2023_cbohf, huang2025} and electronic strong coupling (ESC) \cite{lexander2024, external_field_geometries}. However, the theoretical framework and analytical evaluation of the QED-HF Hessian for polaritons remain unexplored.

Here, we present the derivation and implementation of the analytical QED-HF Hessian and apply it to cavity-molecule systems to investigate how light-matter interactions influence vibrational properties and IR spectra under ESC. 
We choose three prototypical systems to explore different aspects of the cavity effects, starting from a planar system, formaldehyde, a standard benchmark molecule.
$p$-nitroaniline (PNA) is a typical "push-pull" molecule with an electron-donor group NH$_2$ and an acceptor group NO$_2$. Given the high sensitivity of this system due to its large dipole moment, we expect significant cavity-induced modifications in its vibrational properties. Finally, adamantane, with its rigid and spherical-top structure and T$_\text{d}$ symmetry, provides an ideal system for exploring symmetry-breaking effects induced by the cavity field.
As the quantum electromagnetic field modifies the electronic potential energy surface, nuclear motion is consequently altered, leading to changes in the IR spectra. Our results reveal how cavity interactions modify vibrational modes, leading to frequency shifts and intensity modulation in the IR spectra. This paper is organized as follows: in Section \ref{sec:theory}, we derive the QED-HF Hessian expression. 
In Section \ref{sec:validation}, we discuss its implementation and validation. In Section \ref{sec:results}, we report the results for the investigated systems. Finally, in Section \ref{sec:conclusions}, we provide our concluding remarks.

\section{Theory} \label{sec:theory}

\subsection{QED Hamiltonian}

In the Born-Oppenheimer approximation, the interaction between light and matter in the dipole approximation is described by the Pauli-Fierz Hamiltonian, which, in its length-gauge form, is given by \cite{mandal2020polarized,frisk2019ultrastrong}
\begin{gather} \label{eq:H_pf}
    \begin{split}
        H_{\text{PF}} &= H_{\text{M}} + \sum_{\alpha} \omega_{\alpha} b^{\dagger}_{\alpha}b_{\alpha} + \sum_{\alpha} \sum_{pq} \sqrt{\frac{\omega_{\alpha}}{2}} (\boldsymbol{\lambda}_{\alpha} \cdot \mathbf{d})_{pq} (b_{\alpha} + b^{\dagger}_{\alpha})  E_{pq}  \\
        & + \frac{1}{2} \sum_{\alpha} \sum_{pqr} (\boldsymbol{\lambda}_{\alpha} \cdot \mathbf{d})_{pr} (\boldsymbol{\lambda}_{\alpha} \cdot \mathbf{d})_{rq} E_{pq} + \frac{1}{2} \sum_{\alpha} \sum_{pqrs} (\boldsymbol{\lambda}_{\alpha} \cdot \mathbf{d})_{pq} (\boldsymbol{\lambda}_{\alpha} \cdot \mathbf{d})_{rs} e_{pqrs}, 
    \end{split}
\end{gather}
where $H_{\text{M}}$ denotes the molecular Hamiltonian
\begin{gather} \label{eq:H_M}
    H_{\text{M}} = T^{\text{nuc}} + 
    V^{\text{nuc}} + H^{\text{el}} ,
\end{gather}
$T^{\text{nuc}}$ is the nuclear kinetic energy, $V^{\text{nuc}}$ is the nuclear repulsion energy, and $H^{\text{el}}$ is the electronic hamiltonian
\begin{gather} \label{eq:H_e}
    H^{\text{el}} = \sum_{pq} h_{pq} E_{pq} + \frac{1}{2} \sum_{pqrs} g_{pqrs} e_{pqrs}, 
\end{gather}
with $h_{pq}$ and $g_{pqrs}$ denoting the one- and two-electron integrals, respectively. In eqs \ref{eq:H_pf} and \ref{eq:H_e}, the electronic operators are defined as follows
\begin{gather} \label{eq:e_operators}
    \begin{split}
        E_{pq} &= \sum_{\sigma} a^{\dagger}_{p\sigma} a_{q\sigma}, \\ 
        e_{pqrs} &= E_{pq}E_{rs} - \delta_{rq}E_{ps},
    \end{split}
\end{gather}
where the indices $p$, $q$, $r$, and $s$ refer to molecular orbitals, and $a^{\dagger}_{p\sigma}$ and $a_{p\sigma}$ are the creation and annihilation operators for an electron in orbital $p$ with spin $\sigma$. The strong-coupling Hamiltonian in eq \ref{eq:H_pf} includes the quantum electromagnetic field energy, the light-matter interaction explicitly correlating the field and electrons, and the dipole self-energy terms, which ensure the Hamiltonian is bounded from below \cite{rokaj2018light}. Here, the annihilation and creation operators $b_{\alpha}$ and $b^{\dagger}_{\alpha}$ are introduced for photons in mode $\alpha$, with frequency $\omega_{\alpha}$ and polarization vector $\boldsymbol{\lambda}_{\alpha}$
\begin{gather}
    \boldsymbol{\lambda}_{\alpha} = \sqrt{\frac{2 \pi}{\varepsilon_{r} V_{\alpha}}} \boldsymbol{\varepsilon}_{\alpha},
    \label{eq:definition_of_coupling}
\end{gather}
where $\varepsilon_{r}$, $V_{\alpha}$, and $\boldsymbol{\varepsilon}_{\alpha}$ are the relative permittivity, the quantization volume, and the polarization unit vector, respectively. The total dipole moment operator in eq \ref{eq:H_pf} is defined as
\begin{gather} \label{eq:molecular_dipole}
    \sum_{pq} \mathbf{d}_{pq} = \sum_{pq} \left( \mathbf{d}^{\text{el}}_{pq} + \frac{\mathbf{d}^{\text{nuc}}_{pq}}{N_{e}} \delta_{pq} \right) E_{pq},
\end{gather}
where $\mathbf{d}^{\text{el}}_{pq}$ and $\mathbf{d}^{\text{nuc}}_{pq}$ represent the electronic and nuclear dipole moments, respectively, and $N_{e}$ is the number of electrons.

\subsection{QED-HF model}

In the QED-HF model, the wave function is expressed as
\begin{gather}
    \ket{\text{QED-HF}} = \ket{\text{HF}} \otimes \ket{\text{P}},
\end{gather}  
where $\ket{\text{HF}}$ denotes a single Slater determinant, and $\ket{\text{P}}$ is the photon state defined as  
\begin{gather}
    \ket{\text{P}} = \sum_{\mathbf{n}} \prod_{\alpha} \left( b_{\alpha}^{\dagger} \right)^{{n}_{\alpha}} \ket{0} c_{\mathbf{n}},
\end{gather}  
where $\ket{0}$ represents the photonic vacuum state, $c_{\mathbf{n}}$ are the coefficients for the expansion in photon number states, and $\mathbf{n} = (n_{1}, n_{2}, \hdots )$ corresponds to the state with $n_{\alpha}$
photons in mode $\alpha$. The orbitals in the HF reference are optimized using an orthogonal transformation, defined as $\exp(-\kappa)$, where $\kappa$ is an antisymmetric one-electron operator, and the QED-HF energy is minimized with respect to the photon coefficients by diagonalizing the photonic Hamiltonian through an orthogonal coherent state transformation \cite{Tor2020} 
\begin{gather}
    U = \prod_{\alpha} \exp \left(- \frac{\boldsymbol{\lambda}_{\alpha} \cdot \expval{\mathbf{d}}}{\sqrt{2 \omega_{\alpha}}} \left( b_{\alpha} - b_{\alpha}^{\dagger} \right) \right),
\end{gather}  
where  
\begin{gather}
\expval{\mathbf{d}} = \mel{\text{HF}}{\mathbf{d}}{\text{HF}}
\end{gather}  
is the expectation value of the total dipole moment operator $\mathbf{d}$. The transformation of the Hamiltonian in eq \ref{eq:H_pf} gives
\begin{gather} \label{eq:H_pf_coherent}
    \begin{split}
        H & = U^{\dagger} H_{\text{PF}} U \\
        & = H^{\text{el}} + \sum_{\alpha} \omega_{\alpha} b^{\dagger}_{\alpha}b_{\alpha} + \sum_{\alpha} \sum_{pq} \sqrt{\frac{\omega_{\alpha}}{2}} (\boldsymbol{\lambda}_{\alpha} \cdot (\mathbf{d}-\expval{\mathbf{d}} ))_{pq} (b_{\alpha} + b^{\dagger}_{\alpha})  E_{pq}  \\
        & + \frac{1}{2} \sum_{\alpha} \sum_{pqr} (\boldsymbol{\lambda}_{\alpha} \cdot \mathbf{d})_{pr} (\boldsymbol{\lambda}_{\alpha} \cdot \mathbf{d})_{rq} E_{pq} - \sum_{\alpha} \sum_{pq} (\boldsymbol{\lambda}_{\alpha} \cdot \expval{\mathbf{d}}) (\boldsymbol{\lambda}_{\alpha} \cdot \mathbf{d})_{pq} E_{pq} \\
        & + \frac{1}{2} \sum_{\alpha} \sum_{pqrs} (\boldsymbol{\lambda}_{\alpha} \cdot \mathbf{d})_{pq} (\boldsymbol{\lambda}_{\alpha} \cdot \mathbf{d})_{rs} e_{pqrs} + \frac{1}{2} \sum_{\alpha} (\boldsymbol{\lambda}_{\alpha} \cdot \expval{\mathbf{d}})^2.
    \end{split}
\end{gather}
In the coherent-state basis, the reference wave function becomes  
\begin{gather} \label{eq:refwf}
     \ket{\text{R}} = \prod_{\alpha} \exp \left( - \frac{\boldsymbol{\lambda}_{\alpha} \cdot \expval{\mathbf{d}}}{\sqrt{2 \omega_{\alpha}}} \left( b_{\alpha} - b_{\alpha}^{\dagger} \right) \right) \exp \left( -\kappa \right) \ket{\text{QED-HF}},
\end{gather}
and the QED-HF energy is invariant with respect to the choice of molecular origin, even for charged molecules. The energy expression may be written as
\begin{gather}
    E_{\text{QED-HF}} = E_{\text{HF}} + \sum_{\alpha} \sum_{ai} \left( \boldsymbol{\lambda}_{\alpha} \cdot \mathbf{d}_{ai} \right)^2 
\end{gather}
where the labels $i$ and $a$ indicate occupied and virtual orbitals, respectively.

\subsection{Geometrical derivatives of the QED Hamiltonian}

Consider a molecular system in the neighborhood of some reference geometry $\mathbf{R}_{0}$ described by the reference wave function given in eq \ref{eq:refwf}. The corresponding molecular orbitals are expanded in a set of atomic orbitals centered on the nuclei,
\begin{gather} \label{eq:umos}
    \phi_{p} \left( \mathbf{R}_{0} \right) = \sum_{\mu} C_{p \mu} \chi_{\mu} \left( \mathbf{R}_{0} \right).
\end{gather}
This definition can be extended to other molecular geometries by introducing the symmetrically orthonormalized molecular orbitals (OMOs) \cite{helgaker_1984_second,helgaker_molecular_1986, olsen_orbital_1995}
\begin{gather} \label{eq:omos}
    \varphi_p (\mathbf{R}) = \sum_{q} S^{-\frac{1}{2}}_{pq} (\mathbf{R}) \phi_{q} (\mathbf{R})
\end{gather}
where
\begin{gather} \label{eq:s_umo}
    S_{pq} (\mathbf{R}) = \braket{\phi_{p} \left( \mathbf{R} \right) | \phi_{q} \left( \mathbf{R} \right)} = \sum_{\mu \nu} C_{p \mu} C_{q \nu} \braket{\chi_{p} \left( \mathbf{R} \right) | \chi_{q} \left( \mathbf{R} \right)}
\end{gather}
is the overlap between the unmodified molecular orbitals (UMO's) of eq \ref{eq:umos} evaluated at $\mathbf{R}$. By following the theory developed by Helgaker \emph{et al.} in ref \citenum{helgaker_molecular_1986}, we express the Hamiltonian in eq \ref{eq:H_pf_coherent} in the OMO representation as
\begin{gather} \label{eq:H_pf_coherent_omo}
    \begin{split}
        H^{\text{OMO}} \left( \mathbf{R} \right) &= \sum_{pq} \Tilde{h}_{pq}^{\text{OMO}} \left( \mathbf{R} \right) E_{pq} + \frac{1}{2} \sum_{pqrs} \tilde{g}_{pqrs}^{\text{OMO}} \left( \mathbf{R} \right) e_{pqrs} + \sum_{\alpha} \omega_{\alpha} b^{\dagger}_{\alpha}b_{\alpha} \\
        & + \sum_{\alpha} \sum_{pq} \sqrt{\frac{\omega_{\alpha}}{2}} (\boldsymbol{\lambda}_{\alpha} \cdot (\mathbf{d}^{\text{OMO}} \left( \mathbf{R} \right)-\expval{\mathbf{d}^{\text{OMO}}\left( \mathbf{R} \right)} ))_{pq} (b_{\alpha} + b^{\dagger}_{\alpha})  E_{pq}  \\
        & + \frac{1}{2} \sum_{\alpha} (\boldsymbol{\lambda}_{\alpha} \cdot \expval{\mathbf{d}^{\text{OMO}}\left( \mathbf{R} \right)})^2.
    \end{split}
\end{gather}
where the one- and two-electron dipole self-energy contributions are included in the terms $\tilde{h}_{pq}^{\text{OMO}}$ and $\tilde{g}_{pqrs}^{\text{OMO}}$, respectively.
The integrals in eq \ref{eq:H_pf_coherent_omo} can be expressed in the unmodified molecular basis as
\begin{align}
    \Tilde{h}_{pq}^{\text{OMO}} \left( \mathbf{R} \right) &= \sum_{mn} S^{-\frac{1}{2}}_{pm} (\mathbf{R}) S^{-\frac{1}{2}}_{qn} (\mathbf{R}) \Tilde{h}_{mn} \left( \mathbf{R} \right) \label{eq:h_OMO} \\
    \Tilde{g}_{pqrs}^{\text{OMO}} \left( \mathbf{R} \right) &= \sum_{mntu} S^{-\frac{1}{2}}_{pm} (\mathbf{R}) S^{-\frac{1}{2}}_{qn} (\mathbf{R}) S^{-\frac{1}{2}}_{rt} (\mathbf{R}) S^{-\frac{1}{2}}_{su} (\mathbf{R}) \Tilde{g}_{mntu} \left( \mathbf{R} \right) \label{eq:g_OMO} \\
    \mathbf{d}_{pq}^{\text{OMO}} \left( \mathbf{R} \right) &= \sum_{mn} S^{-\frac{1}{2}}_{pm} (\mathbf{R}) S^{-\frac{1}{2}}_{qn} (\mathbf{R}) \mathbf{d}_{mn} \left( \mathbf{R} \right)\label{eq:d_OMO}
\end{align}
and similarly for the dipole integrals and their expectation values. The differentiation of eqs \ref{eq:h_OMO} and \ref{eq:g_OMO} with respect to the coordinate of atom $K$ at the position $\mathbf{R}_{K}$ gives
\begin{align}
    \frac{d \Tilde{h}_{pq}^{\text{OMO}}}{d \mathbf{R}_{K}} &= \frac{d \Tilde{h}_{pq}}{d \mathbf{R}_{K}} - \frac{1}{2} \{ \frac{d S}{d \mathbf{R}_{K}}, \Tilde{h} \}_{pq} \\
    \frac{d \Tilde{g}_{pqrs}^{\text{OMO}}}{d \mathbf{R}_{K}} &= \frac{d \Tilde{g}_{pqrs}}{d \mathbf{R}_{K}} - \frac{1}{2} \{ \frac{d S}{d \mathbf{R}_{K}}, \Tilde{g} \}_{pqrs}
\end{align}
where the one-index transformation terms \cite{helgaker_molecular_1986} are introduced
\begin{align}
    \{ \frac{d S}{d \mathbf{R}_{K}}, \Tilde{h} \}_{pq} &= \sum_{m} ( \frac{d S_{pm}}{d \mathbf{R}_{K}} \tilde{h}_{mq} + \frac{d S_{qm}}{d \mathbf{R}_{K}} \tilde{h}_{mp} ) \\
    \{ \frac{d S}{d \mathbf{R}_{K}}, \Tilde{g} \}_{pqrs} &= \sum_{m} ( \frac{d S_{pm}}{d \mathbf{R}_{K}} \tilde{g}_{mqrs} + \frac{d S_{qm}}{d \mathbf{R}_{K}} \tilde{g}_{pmrs} + \frac{d S_{rm}}{d \mathbf{R}_{K}} \tilde{g}_{pqms} + \frac{d S_{sm}}{d \mathbf{R}_{K}} \tilde{g}_{pqrm} ).
\end{align}
This procedure is generalized to obtain higher derivatives of the Hamiltonian in eq \ref{eq:H_pf_coherent_omo} in terms of UMO-integral derivatives and one-index transformations integrals \cite{helgaker_molecular_1986}, since they are required to obtain Hessian expression.

\subsection{QED-HF Hessian expression} \label{subsec:mol_hess}

To obtain the QED-HF Hessian expression, we first expand the QED-HF electronic energy as \cite{helgaker_1984_second, helgaker_molecular_1986}
\begin{gather} \label{eq:energy_exp}
    \begin{split}
        \Tilde{\mathcal{E}} \left( \mathbf{R} , \Tilde{\boldsymbol{\kappa}} \right)  &= \Tilde{E} \left( \mathbf{R} \right) + \Tilde{\boldsymbol{\kappa}}^{T} \Tilde{\mathbf{g}} \left( \mathbf{R} \right) + \frac{1}{2} \Tilde{\boldsymbol{\kappa}}^{T} \Tilde{\mathbf{F}} \left( \mathbf{R} \right) \Tilde{\boldsymbol{\kappa}} + \mathcal{O} (\Tilde{\kappa}^3),\\
    \end{split}
\end{gather}
where we collected the optimization parameters in vector $\Tilde{\boldsymbol{\kappa}}$. In eq \ref{eq:energy_exp}, we introduced the QED-HF static energy $\Tilde{E} \left( \mathbf{R} \right)$, electronic energy gradient $\Tilde{\mathbf{g}} \left( \mathbf{R} \right)$ and Hessian $\Tilde{\mathbf{F}} \left( \mathbf{R} \right)$. For a fixed geometry $\mathbf{R}$, the optimization condition requires the energy in eq \ref{eq:energy_exp} to be stationary with respect to all orbital variations
\begin{gather} \label{eq:energy_opt_condition}
     \frac{\partial}{\partial \Tilde{\boldsymbol{\kappa}}}   \Tilde{\mathcal{E}} \left( \mathbf{R} , \Tilde{\boldsymbol{\kappa}} \right) = \Tilde{\mathbf{g}} \left( \mathbf{R} \right) + \Tilde{\mathbf{F}} \left( \mathbf{R} \right) \Tilde{\boldsymbol{\kappa}} + \mathcal{O} (\Tilde{\kappa}^2) = \mathbf{0},
\end{gather}
where $\Tilde{\boldsymbol{\kappa}} = \mathbf{0}$ is the solution for $\mathbf{R} = \mathbf{R}_{0}$, with $\mathbf{g} \left(\mathbf{R}_{0} \right) = \mathbf{0}$. By using the condition in eq \ref{eq:energy_opt_condition} we can eliminate the dependence of eq \ref{eq:energy_exp} on the $\Tilde{\boldsymbol{\kappa}}$ parameters and obtain
\begin{gather} \label{eq:opt_energy_geom}
    \begin{split}
        \Tilde{\mathcal{E}} \left( \mathbf{R} \right)  &= \Tilde{E} \left( \mathbf{R} \right) - \frac{1}{2} \Tilde{\boldsymbol{g}}^{T} \left( \mathbf{R} \right) \Tilde{\mathbf{F}}^{-1} \left( \mathbf{R} \right) \Tilde{\boldsymbol{g}} \left( \mathbf{R} \right) + \mathcal{O} (\Tilde{\boldsymbol{g}}^3). 
    \end{split}    
\end{gather}
Here, the geometrical dependence is explicit and confined to the Hamiltonian integrals in eq \ref{eq:H_pf_coherent_omo}. By differentiating eq \ref{eq:opt_energy_geom} at the reference geometry, we obtain the expression of the QED-HF molecular gradient and Hessian
\begin{align}
    \frac{d \Tilde{\mathcal{E}}}{d \mathbf{R}_{K}} &= \frac{d \Tilde{E}}{d \mathbf{R}_{K}} \label{eq:qed_hf_mol_grad} \\
    \frac{d^2 \Tilde{\mathcal{E}}}{d \mathbf{R}_{K} d \mathbf{R}_{L}} &= \frac{d^2 \Tilde{E}}{d \mathbf{R}_{K} d \mathbf{R}_{L}} - \frac{d \Tilde{\mathbf{g}}^{T}}{d \mathbf{R}_{K}}  \mathbf{\Tilde{F}}^{-1} \frac{d \Tilde{\mathbf{g}}}{d \mathbf{R}_{L}}. \label{eq:qed_hf_mol_hess}    
\end{align}
It is worth noting that the QED-HF molecular gradient in eq \ref{eq:qed_hf_mol_grad} only contains a static contribution while the Hessian in eq \ref{eq:qed_hf_mol_hess} has a relaxation contribution due to the response of the electrons to changes in the nuclear coordinates, in accordance with the Born-Oppenheimer approximation. By following the theory outlined in Refs. \citenum{castagnola2023,barlini2024,helgaker_molecular_1986}, we can write the static contribution to the QED-HF Hessian in the UMO basis as
\begin{gather} \label{eq:static_mol_hess}
    \begin{split}
        \frac{d^2 {E}_{\text{QED-HF}}}{d \mathbf{R}_{K} d \mathbf{R}_{L}} &= \sum_{pq} \frac{d^2 \Tilde{F}_{pq} }{d \mathbf{R}_{K} d \mathbf{R}_{L}} \Tilde{\gamma}_{pq} -  \frac{d^2 {S}_{pq}}{d \mathbf{R}_{K} d \mathbf{R}_{L}} \Tilde{F}_{pq} \\
        & -  \frac{d {S}_{pq}}{d \mathbf{R}_{K}} \frac{d {Y}_{pq}}{d \mathbf{R}_{L}} - \frac{d {S}_{pq}}{d \mathbf{R}_{L}} \frac{d {Y}_{pq}}{d \mathbf{R}_{K}} +  \frac{d^2 V^{\text{nuc}}}{d \mathbf{R}_{K} d \mathbf{R}_{L}} \\
        & + \frac{1}{2} \sum_{\alpha} \frac{d^2 (\boldsymbol{\lambda}_{\alpha} \cdot \expval{\mathbf{d}})}{d \mathbf{R}_{K} d \mathbf{R}_{L}},
    \end{split}
\end{gather}
where we introduced the auxiliary matrix 
\begin{gather}
    \frac{d {Y}_{pq}}{d \mathbf{R}_{K}} =  \frac{d \Tilde{F}_{pq}}{d \mathbf{R}_{K}} - \frac{1}{4} \{ \frac{d {S}}{d \mathbf{R}_{K}} , \Tilde{F} \}_{pq} - \frac{1}{2} \sum_{t} \frac{d {S}_{pt}}{d \mathbf{R}_{K}} \Tilde{F}_{tq},
\end{gather}
with $\Tilde{F}_{pq}$ and $\Tilde{\gamma}_{pq}$ being the QED Fock matrix and the QED-HF density matrix, respectively. The relaxation contribution to the Hessian is written as
\begin{gather} \label{eq:relax_mol_hess}
    \begin{split}
        \frac{d \Tilde{\mathbf{g}}^{T}}{d \mathbf{R}_{K}}  \mathbf{\Tilde{F}}^{-1} \frac{d \Tilde{\mathbf{g}}}{d \mathbf{R}_{L}} = \sum_{ai} \mel{\text{QED-HF}}{[ E_{ai}^{-}, (\frac{d {H}}{d \mathbf{R}_{K}})^T ]}{\text{QED-HF}} \frac{d {\kappa}_{ai}}{d \mathbf{R}_{L}},
    \end{split}
\end{gather}
where we used the notation
\begin{gather}
    E_{ai}^{-} = E_{ai} - E_{ia}
\end{gather}
to denote the singlet excitation operator. In eq \ref{eq:relax_mol_hess}, the orbital relaxation to geometric distortions is calculated by solving the linear system of equations
\begin{gather} \label{eq:resp_eq}
    \sum_{bj} \mel{\text{QED-HF}}{[E_{ai},[E_{bj}^{-},H]]}{\text{QED-HF}} \frac{d \kappa_{bj}}{d \mathbf{R}_{K}} = - \mel{\text{QED-HF}}{[ E_{ai}, \frac{d {H}}{d \mathbf{R}_{K}} ]}{\text{QED-HF}}
\end{gather}
for each nucleus and Cartesian component. The above response equations are solved iteratively using a linear subspace solver without explicitly constructing the electronic Hessian on the left-hand side. Note that both the static and relaxation parts of the Hessian in eqs \ref{eq:static_mol_hess} and \ref{eq:relax_mol_hess} are similar to standard HF expressions except for the presence of the dipole self-energy contributions, which are also included in the orbital relaxation of eq \ref{eq:resp_eq}.

\subsection{QED-HF vibrational dipole strength}
In the harmonic approximation, the dipole strength for the $n$-th fundamental vibrational transition is given by \cite{bak_apt}
\begin{gather}
    D_{n} = \frac{1}{2 \omega_{n}} [ {\mathbf{P}}_{n} \cdot {\mathbf{P}}_{n} ]
\end{gather}
where $\omega_{n}$ is the vibrational frequency of the $n$-th transition and ${\mathbf{P}}_{n}$ is the polar tensor in normal coordinates
\begin{gather}
    {\mathbf{P}}_{n} = \sum_{K} \mathbf{P}_{K} \mathbf{L}_{K,n},
\end{gather}
where $\mathbf{L}_{K,n}$ is the Cartesian-to-normal coordinates transformation matrix and $\mathbf{P}_{K}$ are the components of the atomic polar tensor (APT) with respect to the coordinates of the $K$-th nucleus
\begin{gather} \label{eq:apt}
    \mathbf{P}_{K} = \mathbf{E}_{K} + \mathbf{N}_{K}
\end{gather}
where $\mathbf{E}_{K}$ and $\mathbf{N}_{K}$ denote the electronic and nuclear contributions, respectively. As discussed in ref \citenum{schnappinger2023}, and building on the formalism in Sec. \ref{subsec:mol_hess}, the electronic contribution to the APT for polaritonic systems is expressed as
\begin{gather} \label{eq:apt_el}
    \mathbf{E}_{K} = \sum_{pq} (\frac{d \mathbf{d}^{\text{el}}_{pq}}{d \mathbf{R}_{K}} - \frac{1}{2} \{ \frac{d S}{d \mathbf{R}_{K}}, \mathbf{d}^{\text{el}} \}_{pq} ) \tilde{\gamma}_{pq} - 4 \sum_{ai} \frac{d {\kappa}_{ai}}{d \mathbf{R}_{K}} \mathbf{d}^{\text{el}}_{ai}  
\end{gather}
where the orbital relaxation to geometric distortions is required. It is important to note that the QED contribution in eq \ref{eq:apt_el} is included through both the density matrix and the orbital relaxation term. The pure nuclear contribution in eq \ref{eq:apt} can be expressed as
\begin{gather} 
    \mathbf{N}_{K} = Z_{K} \mathbf{1} \label{eq:apt_nuc},
\end{gather}
where $Z_{K}$ is the nuclear charge of the nucleus $K$ and $\mathbf{1}$ is the identity matrix.

\section{Validation and implementation} \label{sec:validation}

The calculation of the QED-HF Hessian, vibrational frequencies, and dipole strengths has been implemented in a development version of the e$^{\mathcal{T}}$ program \cite{folkestad2020t}. This implementation follows the standard procedure for the calculation of vibrational properties \cite{helgaker_molecular_1986}. The QED-HF Hessian code was validated by numerical differentiation of the QED-HF molecular gradient. The geometric derivatives of the integrals were calculated using phasedINT \cite{phasedint2025}, a library for computing Gaussian integrals in computational chemistry. The QED-HF vibrational dipole strength was validated by setting the coupling strength to zero and comparing the results with those obtained from the corresponding HF code.

\section{Results and discussions} \label{sec:results}

The molecular geometries were optimized at the QED-HF level using the aug-cc-pVDZ basis set at the different coupling strengths. The code utilizes an analytical gradient and translational-rotational internal coordinates (TRIC) \cite{Lee-Ping_2016_TRIC}, allowing for the inclusion of both internal coordinates and the rotational degrees of freedom required to account for the quantum electromagnetic field. In this way, we optimized the molecular orientation with respect to the cavity field by considering a single orientation, thus avoiding computationally expensive rotational averaging. This approach provides a lower bound for cavity-induced effects since other orientations lead to a stronger interaction, since the dipole self-energy contribution is not minimized. The IR spectra shown below were obtained using Lorentzian distribution functions with a half width at half maximum (HWHM) of 10 cm$^{-1}$. The intensities are normalized with respect to the intensity of the most intense peak overall.
For all calculations, we employed coupling values of 0.00, 0.05, and 0.10 a.u. These relatively large couplings were chosen to better highlight the effects, even though such values are currently experimentally unfeasible.

\subsection{Formaldehyde}

\begin{figure}[H]
    \centering
    \includegraphics[width=0.75\linewidth]{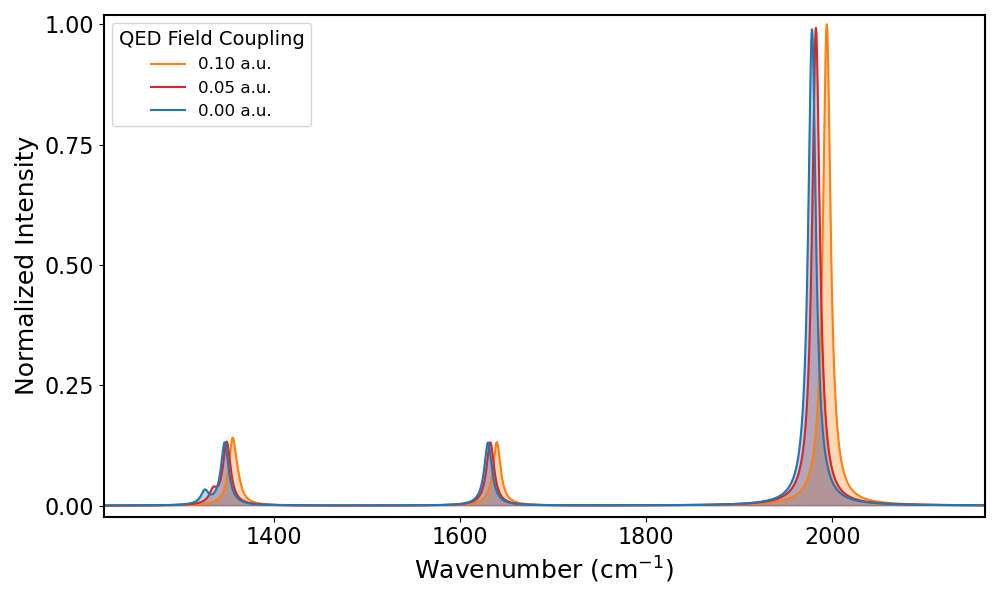}
    \caption{Fingerprint region of the IR spectrum of formaldehyde obtained for coupling values of 0.00, 0.05, and 0.10 a.u.}
    \label{fig:HCHO_1200_2000}
\end{figure}

\begin{figure}[H]
    \centering
    \includegraphics[width=0.75\linewidth]{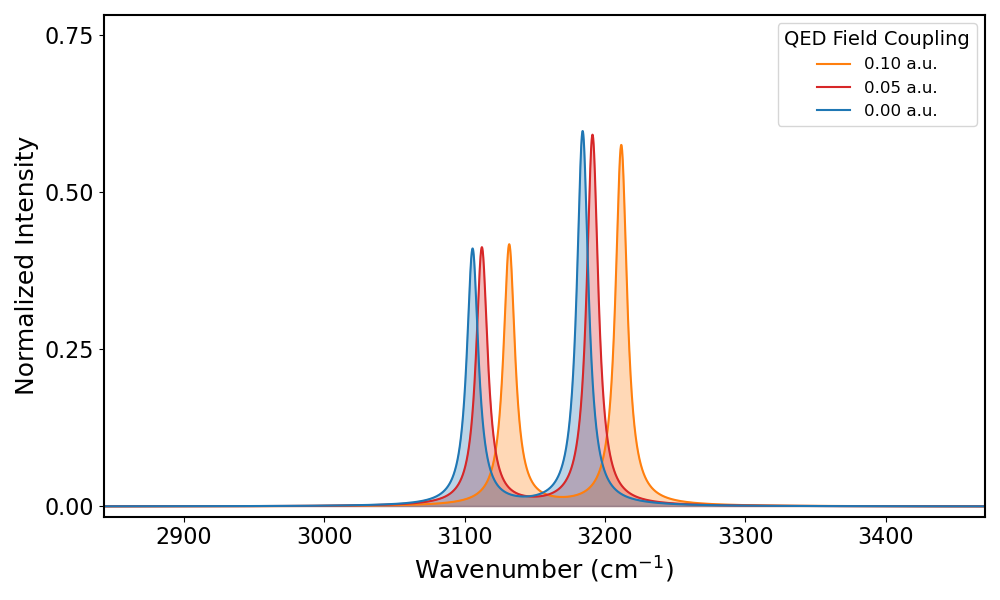}
    \caption{C-H stretching region of the formaldehyde IR spectrum obtained for coupling values of 0.00, 0.05, and 0.10 a.u.}
    \label{fig:HCHO_2900_3400}
\end{figure}

The IR spectra of formaldehyde for coupling values of 0.00, 0.05, and 0.10 a.u. are reported in Figure \ref{fig:HCHO_1200_2000} for the fingerprint region and Figure \ref{fig:HCHO_2900_3400} for the CH-stretching region, while the vibrational frequencies and the integrated intensities are listed in Table~\ref{tab:HCHO_ir_total}.
The spectrum obtained in the absence of the quantum electromagnetic field is qualitatively in agreement with experimental data\cite{nist_webbook}, as shown for instance in ref \citenum{leszczynski1997}

\begin{figure}[H]
    \centering
    \includegraphics[width=0.75\linewidth]{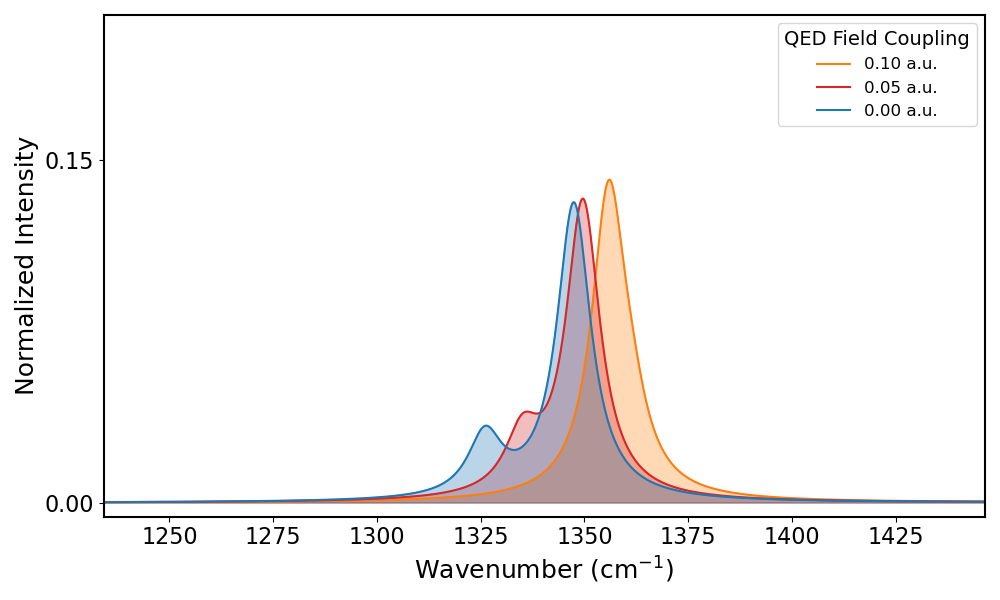}
    \caption{Region of the formaldehyde IR spectrum containing the wagging and rocking normal modes for coupling values of 0.00, 0.05, and 0.10 a.u.}
    \label{fig:HCHO_1200}
\end{figure}

\begin{figure}[H]
    \centering
    \includegraphics[width=0.85\linewidth]{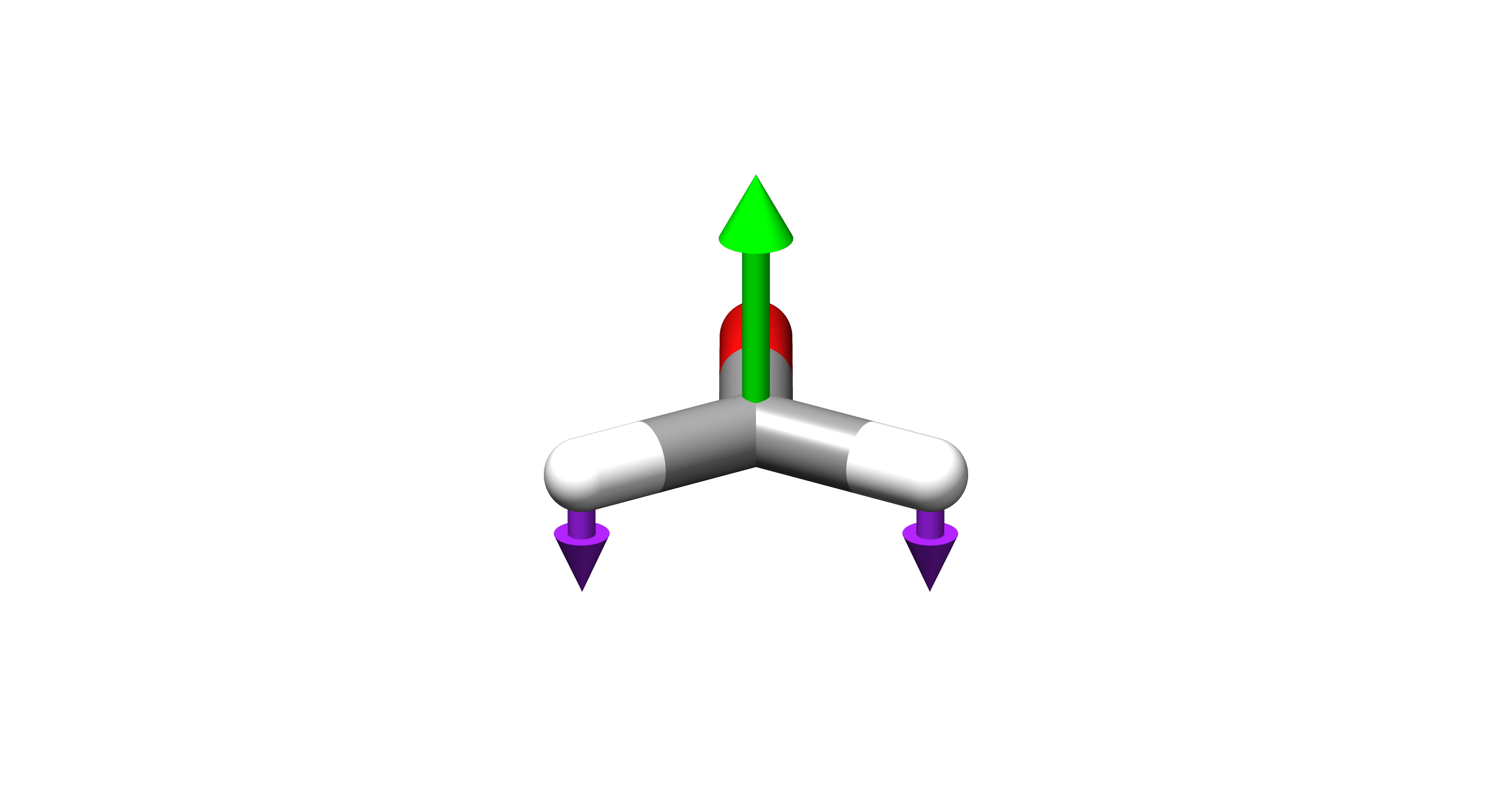}
    \caption{Wagging normal mode in formaldehyde. Purple arrows represent the nuclear displacements, and the green arrow the polarization orientation of the quantum electromagnetic field.}
    \label{fig:HCHO_normal_mode_7}
\end{figure}

As the coupling increases, we observe a blue shift for all bands, which is expected to be attributed to the contraction of the electronic density induced by the quantum electromagnetic field. Interestingly, as the coupling increases, the wagging vibrational mode becomes more destabilized compared to the rocking mode. 
At a coupling value of 0.10 a.u., the wagging mode becomes higher in energy than the rocking mode, as shown in Table~\ref{tab:HCHO_ir_total}.
Because of the proximity of their energies they appear as a single band centered at about 1355 cm$^{-1}$, as shown in Figure~\ref{fig:HCHO_1200}. 
This behavior can be explained by the effect of the cavity on the electronic density. In the case of formaldehyde, the most stable configuration is the one in which the molecular dipole moment is orthogonal to the polarization of the electromagnetic field in the cavity. 
In this situation, the cavity confines the electronic density onto the molecular plane. \cite{barlini2024}
As a result, vibrational modes that involve nuclear motions out of the molecular plane are inhibited, as in the case of wagging, depicted in Figure \ref{fig:HCHO_normal_mode_7}. 
Another consequence is a reduction of the intensity as the nuclear displacements are restrained. Regarding the other normal modes, we observe an increase in intensity for the C=O stretching and the symmetric C-H stretching, and a decrease in intensity for the antisymmetric C-H stretching mode. Moreover, in Table~\ref{tab:HCHO_ir_total} we observe two new low-frequency signals. 
The first corresponds to the rotation of formaldehyde around an axis orthogonal to both the polarization direction and the molecular dipole moment. 
The second also arises from a rotational mode.
However, its intensity is null, as the associated rotation occurs around the molecular symmetry axis and does not induce a change in the dipole moment.

\begin{table}[H]
\centering
\begin{tabular}{lrrrrrr}
\hline
  & \multicolumn{2}{c}{0.00 } & \multicolumn{2}{c}{0.05 } & \multicolumn{2}{c}{0.10 } \\
\hline
Assignment & {$\tilde{\nu}$} & {$\mathcal{I}$} & {$\tilde{\nu}$} & {$\mathcal{I}$} & {$\tilde{\nu}$} & {$\mathcal{I}$} \\
\hline
rotation &  - & - & 54.0  & 25.73  & 106.3  & 26.80  \\
rotation &  - & - & 75.1  & 0.00  & 148.0  & 0.00  \\
wagging & 1326.1 & 4.24  & 1335.1 & 3.98  & 1361.2 & 3.31  \\
rocking & 1347.5 & 20.45 & 1349.7 & 20.50  & 1355.8 & 20.67  \\
scissoring & 1630.4 & 20.61 & 1632.8 & 20.64  & 1639.7 & 20.74  \\
C=O stretching & 1978.3 & 155.61 & 1982.5 & 156.07 & 1994.2 & 157.26 \\
C-H stretching (sym.) & 3105.7 & 64.21 & 3112.3 & 64.53  & 3131.8 & 65.29  \\
C-H stretching (asym.) & 3184.1 & 93.73 & 3191.2 & 92.82  & 3211.7 & 90.23  \\
\hline
\end{tabular}
\caption{Vibrational energies (cm$^{-1}$) and IR intensities (km mol$^{-1}$) of formaldehyde at different coupling strengths (a.u.).}
\label{tab:HCHO_ir_total}
\end{table}

\subsection{$p$-nitroaniline}

\begin{figure}[H]
    \centering
    \includegraphics[width=0.75\linewidth]{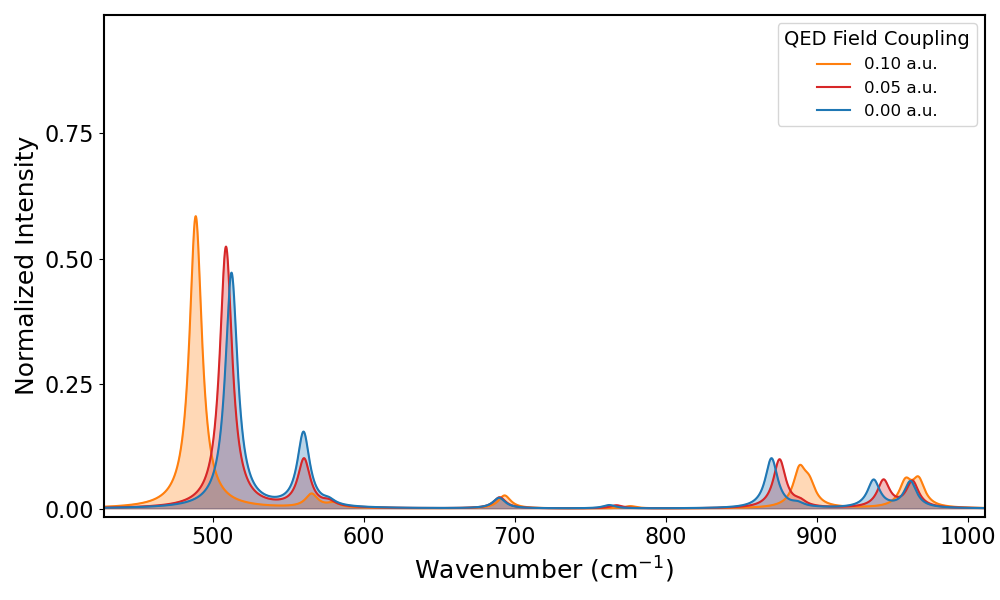}
    \caption{IR spectra of PNA from 450 to 1000 cm$^{-1}$, obtained for coupling values of 0.00, 0.05, and 0.10 a.u.}
    \label{fig:PNA_500_100}
\end{figure}

\begin{figure}[H]
    \centering
    \includegraphics[width=0.75\linewidth]{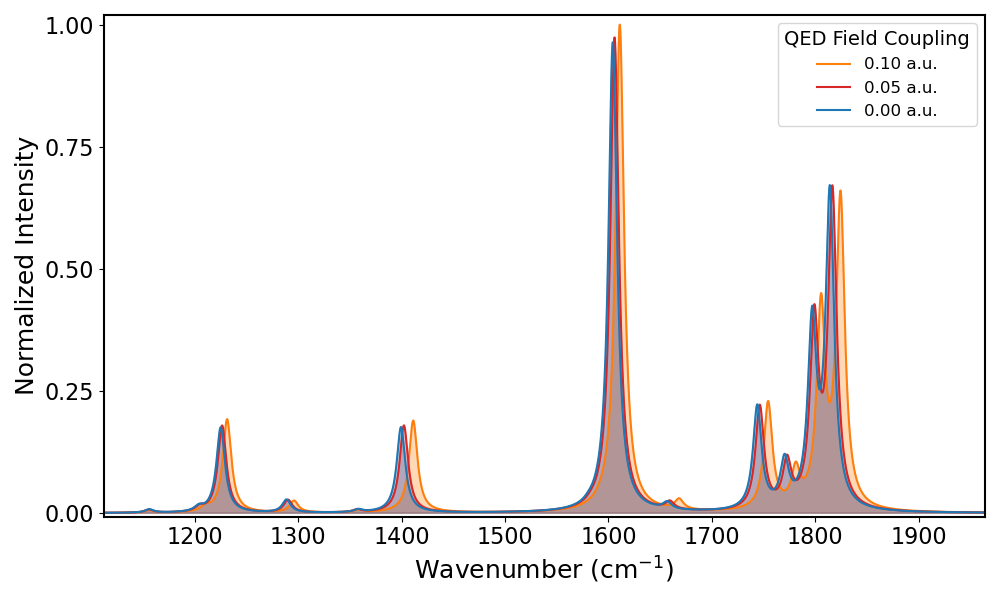}
    \caption{IR spectra of PNA in the 1150 to 1900 cm$^{-1}$ region obtained for coupling values of 0.00, 0.05, and 0.10 a.u.}
    \label{fig:PNA_1200_2000}
\end{figure}

\begin{figure}[H]
    \centering
    \includegraphics[width=0.75\linewidth]{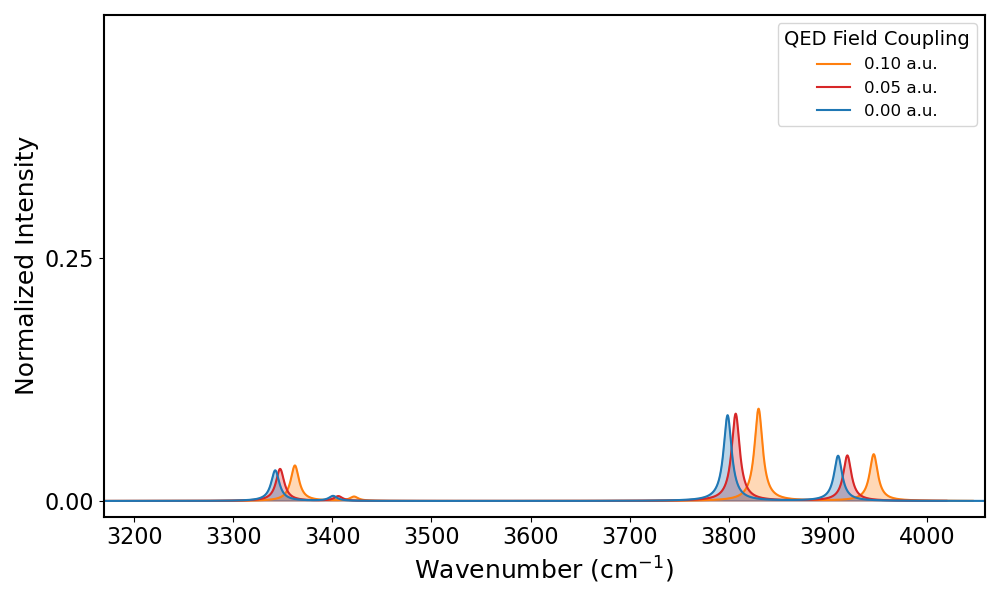}
    \caption{IR spectra of PNA in the 3200 to 4000 cm$^{-1}$ region obtained for coupling values of 0.00, 0.05, and 0.10 a.u.}
    \label{fig:PNA_3200_4000}
\end{figure}

\begin{figure}[H]
    \begin{subfigure}{0.5\textwidth}
    \caption{}
    \centering
    \includegraphics[width=\textwidth]{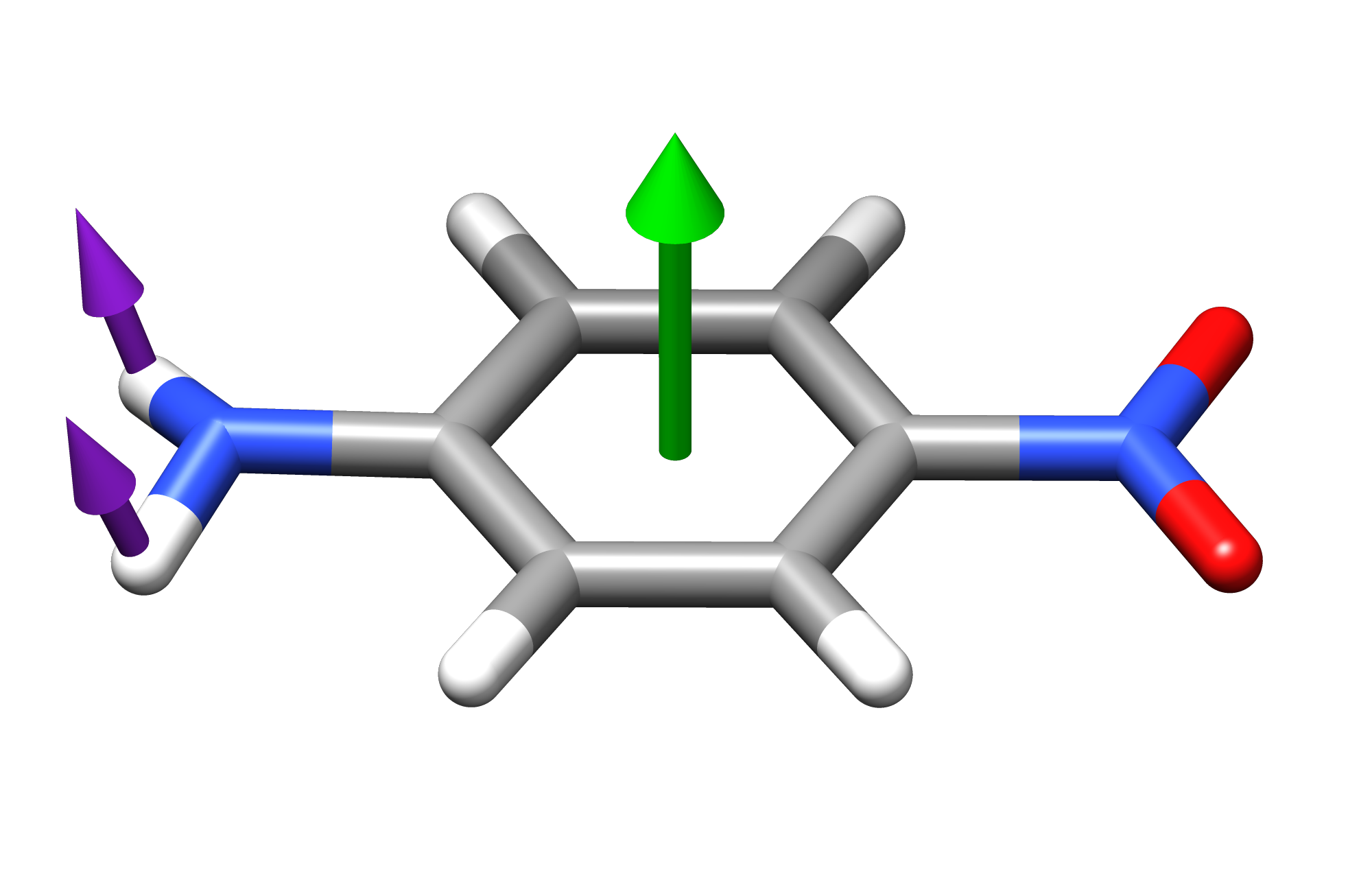}
    \label{fig:PNA_normal_mode_15}
    \end{subfigure}%
    \begin{subfigure}{0.5\textwidth}
    \caption{}
    \centering
    \includegraphics[width=\textwidth]{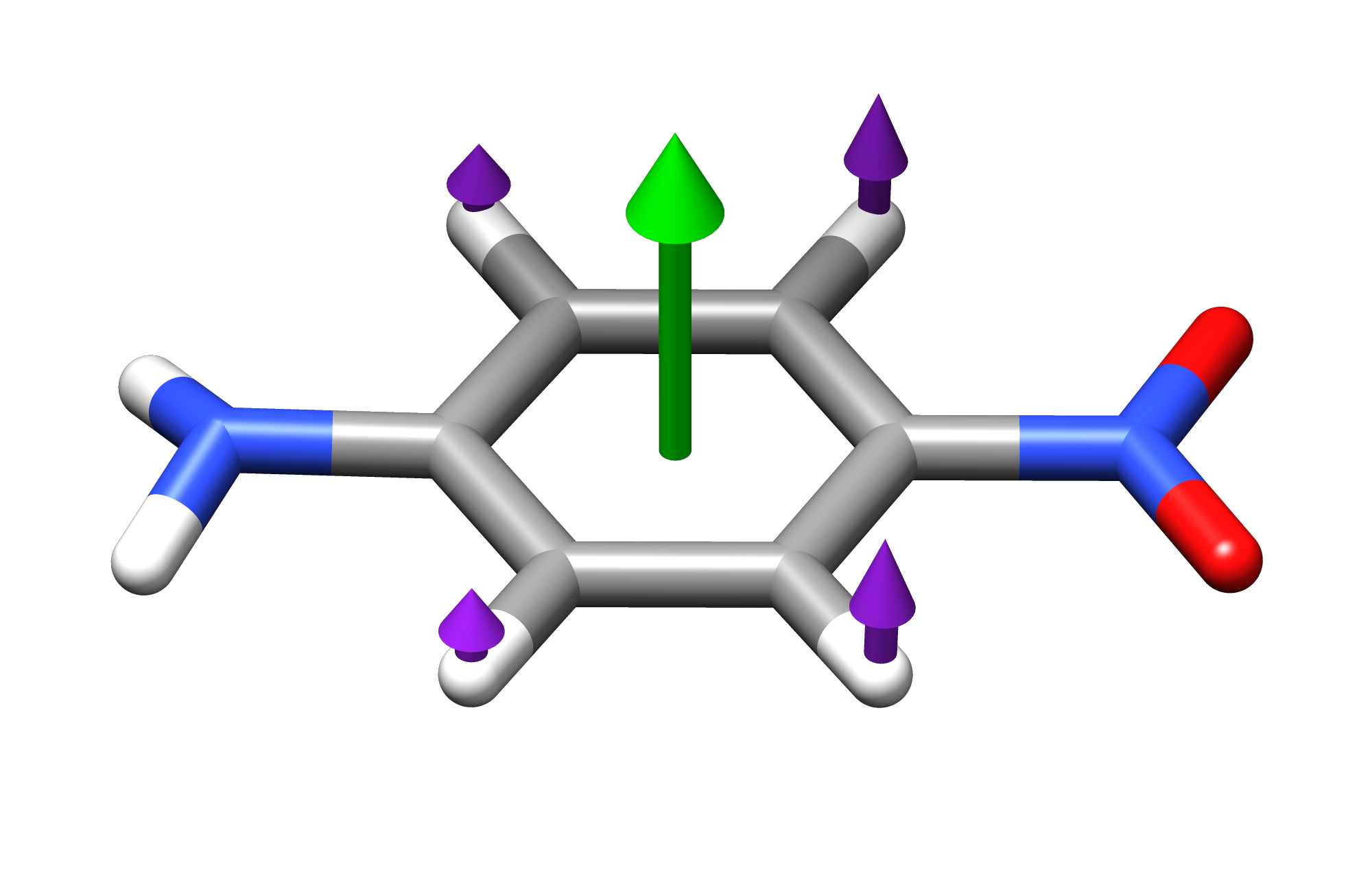}
    \label{fig:PNA_normal_mode_16}
    \end{subfigure}
    \caption{NH$_2$ rocking (a) and out-of-plane molecular wagging mode (b) in PNA. Purple arrows show the nuclear displacements, while green arrows represent the polarization orientation of the quantum electromagnetic field.}
\end{figure}

\begin{figure}[H]
    \begin{subfigure}{0.5\textwidth}
    \caption{}
    \centering
    \includegraphics[width=\textwidth]{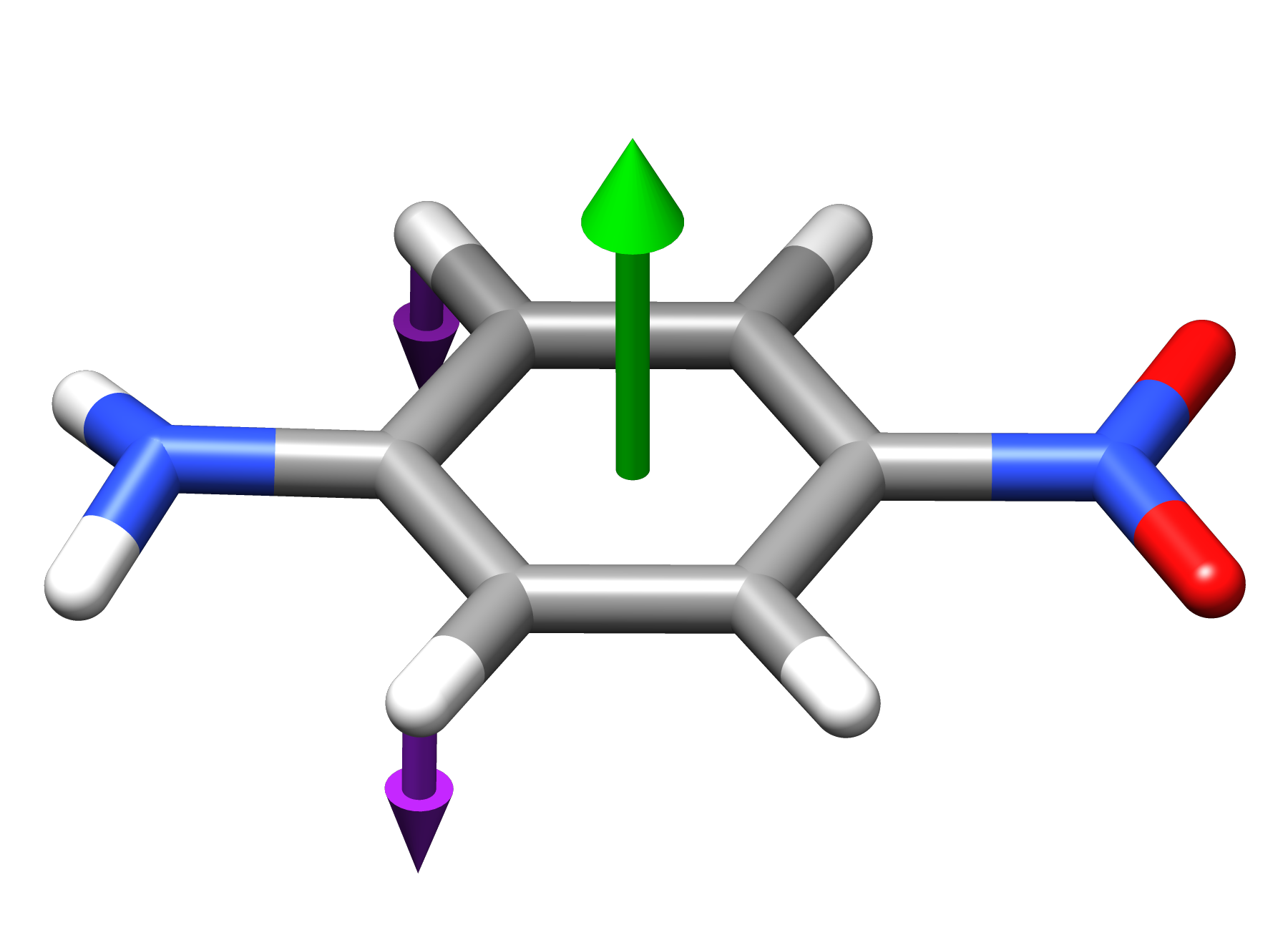}
    \label{fig:PNA_normal_mode_24}
    \end{subfigure}%
    \begin{subfigure}{0.5\textwidth}
    \caption{}
    \centering
    \includegraphics[width=\textwidth]{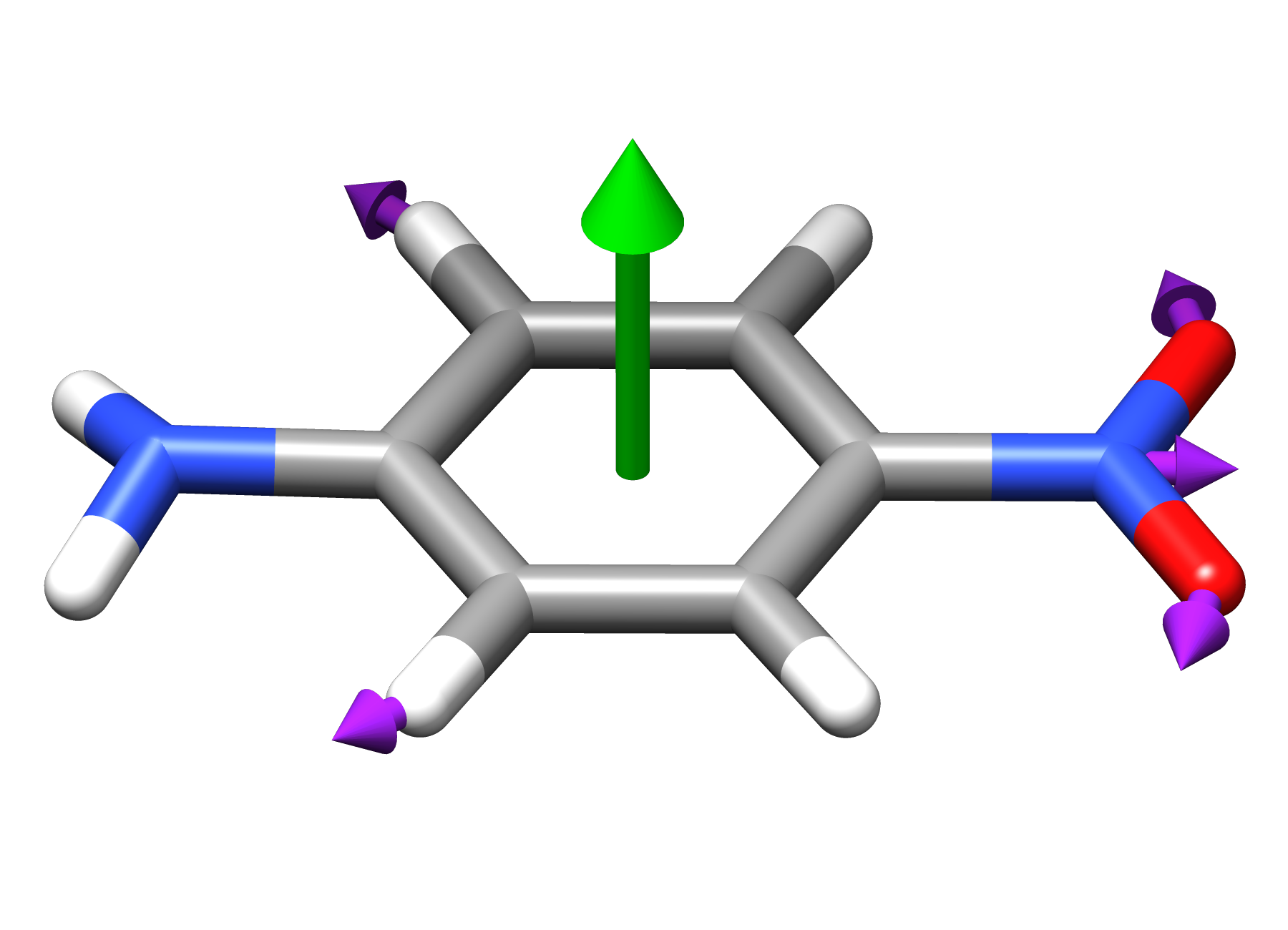}
    \label{fig:PNA_normal_mode_25}
    \end{subfigure}
    \caption{Normal modes associated with the transitions at 938 cm$^{-1}$ (a) and 962 cm$^{-1}$ (b) in PNA. Purple arrows show the nuclear displacements, while green arrows represent the polarization orientation of the quantum electromagnetic field.}
\end{figure}

The IR spectrum of PNA at different coupling values of 0.00, 0.05, and 0.10 a.u. is shown in Figure~\ref{fig:PNA_500_100} from 450 to 1100 cm$^{-1}$, Figure~\ref{fig:PNA_1200_2000} from 1150 to 1900 cm$^{-1}$, and Figure~\ref{fig:PNA_3200_4000} from 3200 to 4000 cm$^{-1}$. 
The corresponding vibrational energies and IR intensities are reported in Table~\ref{tab:PNA_ir_total}.
Comparison with experimental data in absence of the cavity field shows qualitative agreement. \cite{nist_webbook}
In Figure~\ref{fig:PNA_500_100}, the band around 500~cm$^{-1}$ corresponds to a wagging mode of the amino group, shown in Figure \ref{fig:PNA_normal_mode_15}. 
As the coupling increases it becomes red-shifted while its intensity increases. 
This wagging mode corresponds to a nuclear displacement that drives the molecule toward a planar geometry. 
In the presence of a quantum electromagnetic field, the cavity interaction stabilizes this planar configuration, leading to a reduction in the vibrational energy. 
In such a configuration, the variation of the dipole moment with respect to the normal mode increases, leading to an enhancement in the intensity. 
In contrast, the out-of-plane wagging mode at 560 cm$^{-1}$, shown in Figure \ref{fig:PNA_normal_mode_16}, is suppressed since it involves further distorsion with the displacement of the hydrogen atoms in the amino group away from the molecular plane, resulting in an increase of the frequency and a decrease in the intensity at higher couplings.
Considering the modes related to the in- and out-of-plane deformations of the benzene moiety, whose vibrational energies are around 950 cm$^{-1}$, they exhibit a behavior similar to what observed for formaldehyde.
The lower-frequency one in Figure \ref{fig:PNA_normal_mode_24} exhibits a larger shift than the other one in Figure \ref{fig:PNA_normal_mode_25}. The former involves significant out-of-plane nuclear displacements, strongly restricted by the quantum electromagnetic field, whereas the latter, characterized by in-plane displacements, experiences a smaller frequency shift.
Similar behavior can be observed in other regions of the IR spectrum at higher vibrational energies. This behavior could be attributed to bond strengthening induced by electron density contraction caused by the quantum electromagnetic field. Again, this result is in line with what was previously noted with formaldehyde.

\begin{table}[H]
\centering
\begin{tabular}{rrrrrr}
\hline
 \multicolumn{2}{c}{0.00 } & \multicolumn{2}{c}{0.05 } & \multicolumn{2}{c}{0.10 } \\
\hline
{$\tilde{\nu}$} & {$\mathcal{I}$} & {$\tilde{\nu}$} & {$\mathcal{I}$} & {$\tilde{\nu}$} & {$\mathcal{I}$} \\
\hline
512.5  & 313.02  & 508.8  & 348.18  & 488.8  & 389.18  \\
560.1  & 98.71   & 560.5  & 63.51   & 565.5  & 18.00   \\
576.9  & 5.76    & 576.9  & 5.78    & 580.0  & 5.80    \\
689.7  & 11.93   & 689.6  & 1.70    & 693.2  & 15.68   \\
689.7  & 2.09    & 690.1  & 13.10   & 694.0  & 1.65    \\
762.6  & 4.42    & 767.3  & 4.14    & 776.5  & 3.07    \\
870.0  & 66.43   & 875.4  & 64.64   & 888.6  & 47.63   \\
887.5  & 4.50    & 888.9  & 6.36    & 895.1  & 26.42   \\
902.0  & 0.02    & 908.7  & 0.02    & 927.3  & 0.03    \\
937.7  & 36.95   & 944.2  & 36.31   & 958.8  & 31.97   \\
962.1  & 35.07   & 963.4  & 34.91   & 967.5  & 34.34   \\
1070.0 & 1.19    & 1077.6 & 1.18    & 1091.0 & 1.65    \\
1089.6 & 0.58    & 1090.5 & 0.64    & 1095.6 & 0.92    \\
1093.7 & 0.01    & 1102.0 & 0.01    & 1119.2 & 0.01    \\
1156.0 & 4.65    & 1155.8 & 4.26    & 1156.0 & 3.38    \\
1203.9 & 6.27    & 1204.9 & 6.30    & 1210.6 & 6.50    \\
1225.1 & 116.31  & 1226.5 & 118.81  & 1231.3 & 127.37  \\
1288.7 & 17.50   & 1290.1 & 17.25   & 1296.0 & 16.03   \\
1358.3 & 3.70    & 1357.4 & 3.68    & 1358.9 & 3.56    \\
1399.4 & 116.73  & 1402.3 & 118.99  & 1411.3 & 125.60  \\
1434.2 & 0.09    & 1436.0 & 0.08    & 1442.8 & 0.06    \\
1580.7 & 2.05    & 1583.0 & 1.96    & 1591.5 & 2.03    \\
1604.2 & 640.96  & 1606.0 & 647.58  & 1611.0 & 664.93  \\
1657.0 & 9.52    & 1659.4 & 10.48   & 1668.4 & 13.93   \\
1744.0 & 141.11  & 1746.4 & 140.61  & 1754.5 & 145.75  \\
1770.4 & 61.67   & 1772.7 & 60.33   & 1781.2 & 48.53   \\
1797.0 & 245.46  & 1799.1 & 248.74  & 1805.6 & 268.42  \\
1814.2 & 426.00  & 1816.6 & 426.26  & 1824.5 & 420.41  \\
3342.0 & 9.30    & 3346.9 & 9.90    & 3361.7 & 11.32   \\
3342.7 & 11.73   & 3347.6 & 12.15   & 3362.6 & 13.16   \\
3400.7 & 0.22    & 3406.0 & 0.37    & 3421.9 & 0.103    \\
3400.7 & 3.13    & 3406.1 & 2.81    & 3421.9 & 2.73    \\
3798.8 & 58.63   & 3807.0 & 59.68   & 3830.1 & 63.09   \\
3910.3 & 30.85   & 3919.7 & 31.06   & 3946.3 & 31.89   \\
\hline
\end{tabular}
\caption{Vibrational energies (cm$^{-1}$) and IR intensities (km mol$^{-1}$) of PNA in the absence and presence of the quantum electromagnetic field at different couplings (a.u.).}
\label{tab:PNA_ir_total}
\end{table}

\subsection{Adamantane}

\begin{figure}[H]
    \centering
    \includegraphics[width=0.75\linewidth]{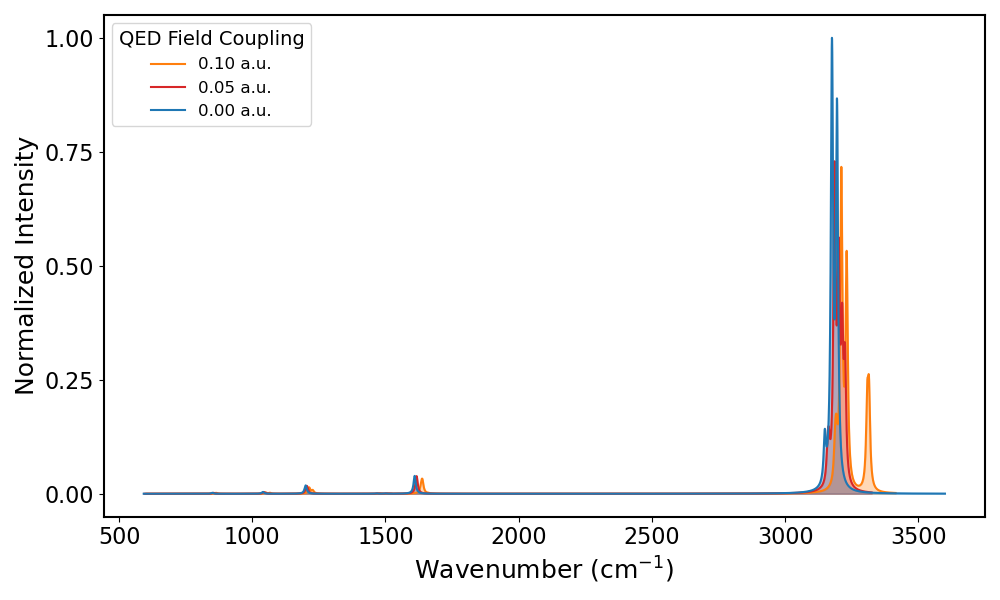}
    \caption{IR spectra of adamantane obtained for coupling values of 0.00, 0.05, and 0.10 a.u.}
    \label{fig:ADA}
\end{figure}

As final example, the IR spectra of adamantane are shown in Figure~\ref{fig:ADA}, while the predicted vibrational energies and IR intensities corresponding to allowed transitions in absence of the quantum electromagnetic field are collected in Table~\ref{tab:ADA_ir}.
These results show qualitative agreement with previous investigations \cite{lahorija2002} and experimental data \cite{nist_webbook}.

\begin{figure}[H]
    \centering
    \includegraphics[width=0.75\linewidth]{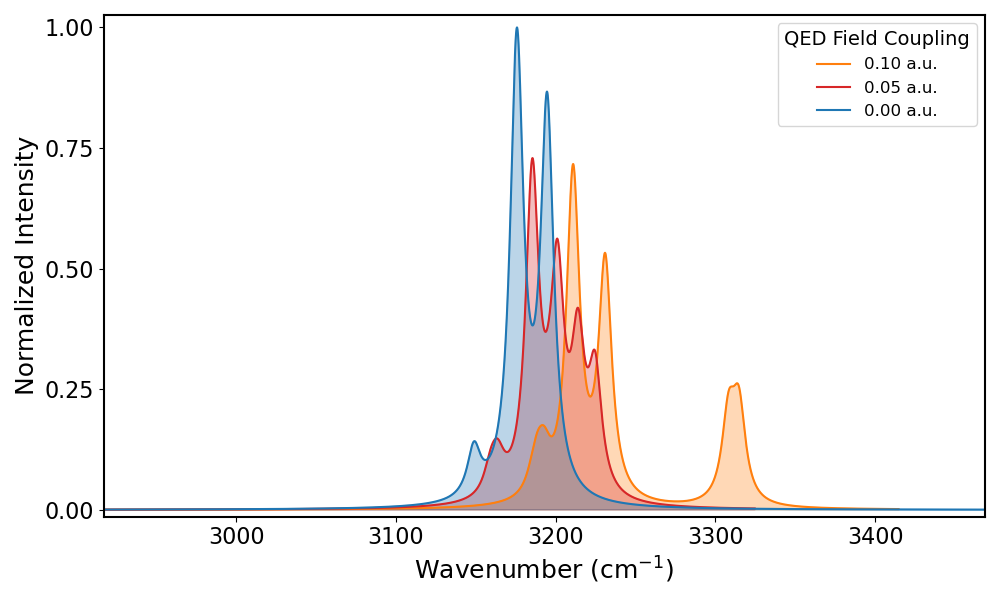}
    \caption{IR spectra in the CH-stretching region of adamantane obtained for coupling values of 0.00, 0.05, and 0.10 a.u.}
    \label{fig:ADA_3100_3400}
\end{figure}

\begin{table}[H]
\centering
\begin{tabular}{crr}
\hline
{Irrep} & {$\tilde{\nu}$ } & {$\mathcal{I}$ } \\
\hline
    T$_2$ & 850.0 & 0.39 \\
    T$_2$ & 1200.1 & 3.51 \\
    T$_2$ & 1609.5 & 7.50 \\
    T$_2$ & 3175.7 & 181.05 \\
    T$_2$ & 3148.8 & 19.37 \\
    T$_2$ & 3194.7 & 154.20 \\
\hline
\end{tabular}
\caption{Vibrational energies (cm$^{-1}$) and IR intensities (km mol$^{-1}$) for non-null fundamental transitions of adamantane in the absence of the quantum electromagnetic field.}
\label{tab:ADA_ir}
\end{table}
In the absence of the cavity field, the molecule has a T$_\text{d}$ symmetry and only transitions to states of T$_2$ symmetry are IR active.
Once the cavity field is applied, the preferred orientation corresponds to the polarization aligned along one of the C$_3$ axes. 
This causes a slight contraction along that axis and a lowering of the symmetry to the C$_{3v}$ group.
The allowed transitions are now those leading to a final state of A$_1$ or E symmetry. 
Due to the cavity field-driven symmetry breaking, the T$_1$ and T$_2$ transitions are split into A$_2$ and E, and A$_1$ and E transitions, respectively. The vibrational energies with non-null IR intensities are reported in Tables~\ref{tab:ADA_ir_0.05} and~\ref{tab:ADA_ir_0.10} for coupling values of 0.05 and 0.10 a.u. respectively, and the IR spectrum in the C-H stretching region is shown in Figure \ref{fig:ADA_3100_3400}.

\begin{table}[H]
\centering
\begin{tabular}{crr}
\hline
{Irrep} & {$\tilde{\nu}$ } & {$\mathcal{I}$ } \\
\hline
    A$_1$ & 853.1 & 0.41 \\
    E & 854.0 & 0.39 \\
    E & 1203.7 & 3.56 \\
    A$_1$ & 1207.0 & 3.66 \\
    A$_1$ & 1615.9 & 7.78 \\
    E & 1617.0 & 7.33 \\
    E & 3158.8 & 10.20 \\
    A$_1$ & 3160.7 & 11.92 \\
    A$_1$ & 3163.5 & 33.06 \\
    E & 3165.7 & 3.46 \\
    E & 3185.4 & 191.02 \\
    A$_1$ & 3195.9 & 21.60 \\
    E & 3201.1 & 124.17 \\
    E & 3213.7 & 23.83 \\
    A$_1$ & 3214.1 & 122.20 \\
    A$_1$ & 3224.7 & 141.25 \\
\hline
\end{tabular}
\caption{Vibrational energies (cm$^{-1}$) and IR intensities (km mol$^{-1}$) of adamantane with a coupling of 0.05 a.u.}
\label{tab:ADA_ir_0.05}
\end{table}

\begin{figure}[H]
    \centering
    \begin{subfigure}{0.4\textwidth}
    \caption{}
    \centering
    \includegraphics[width=\textwidth]{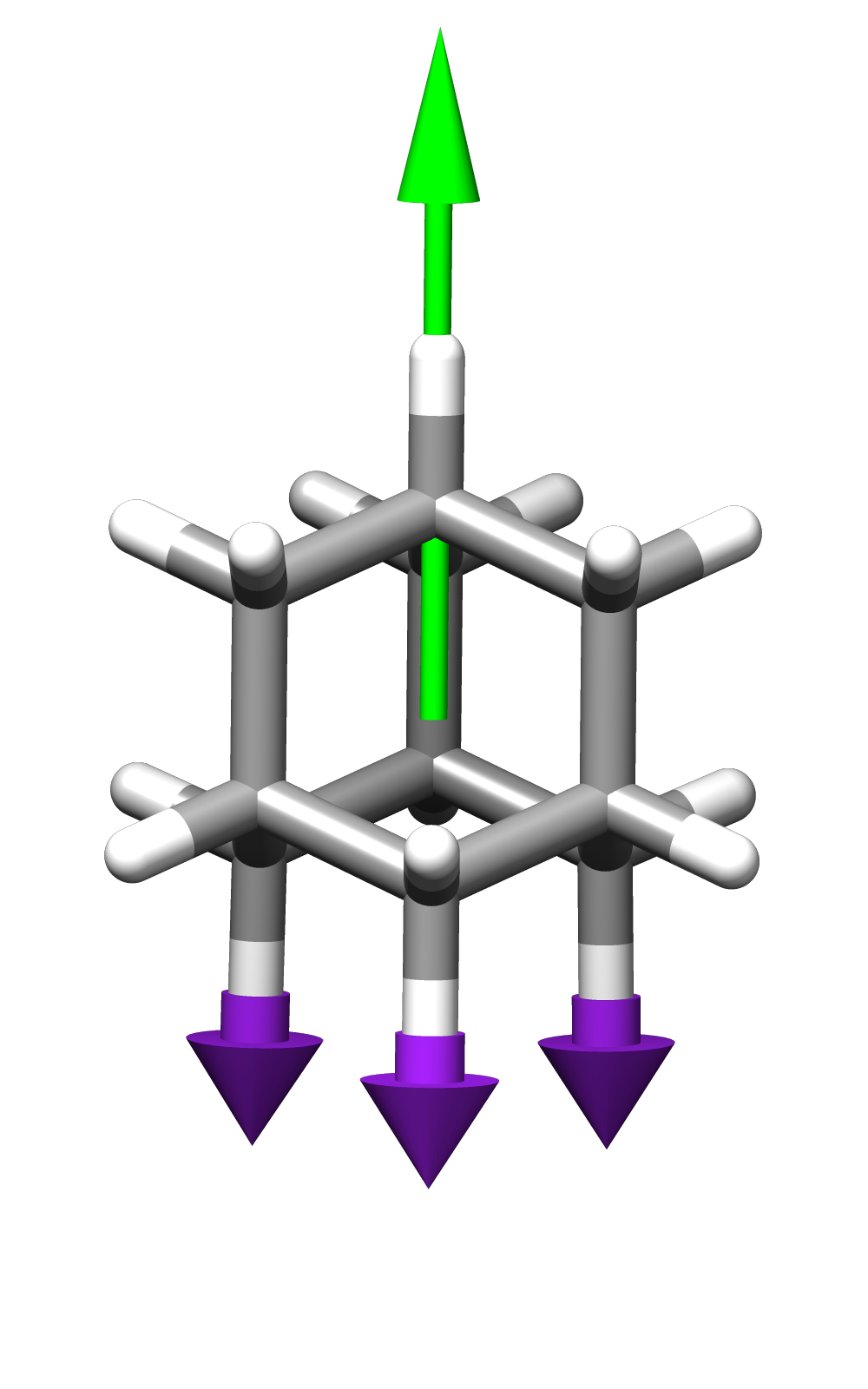}
    \label{fig:ADA_normal_mode_78}
    \end{subfigure}
    \begin{subfigure}{0.4\textwidth}
    \caption{}
    \centering
    \includegraphics[width=\textwidth]{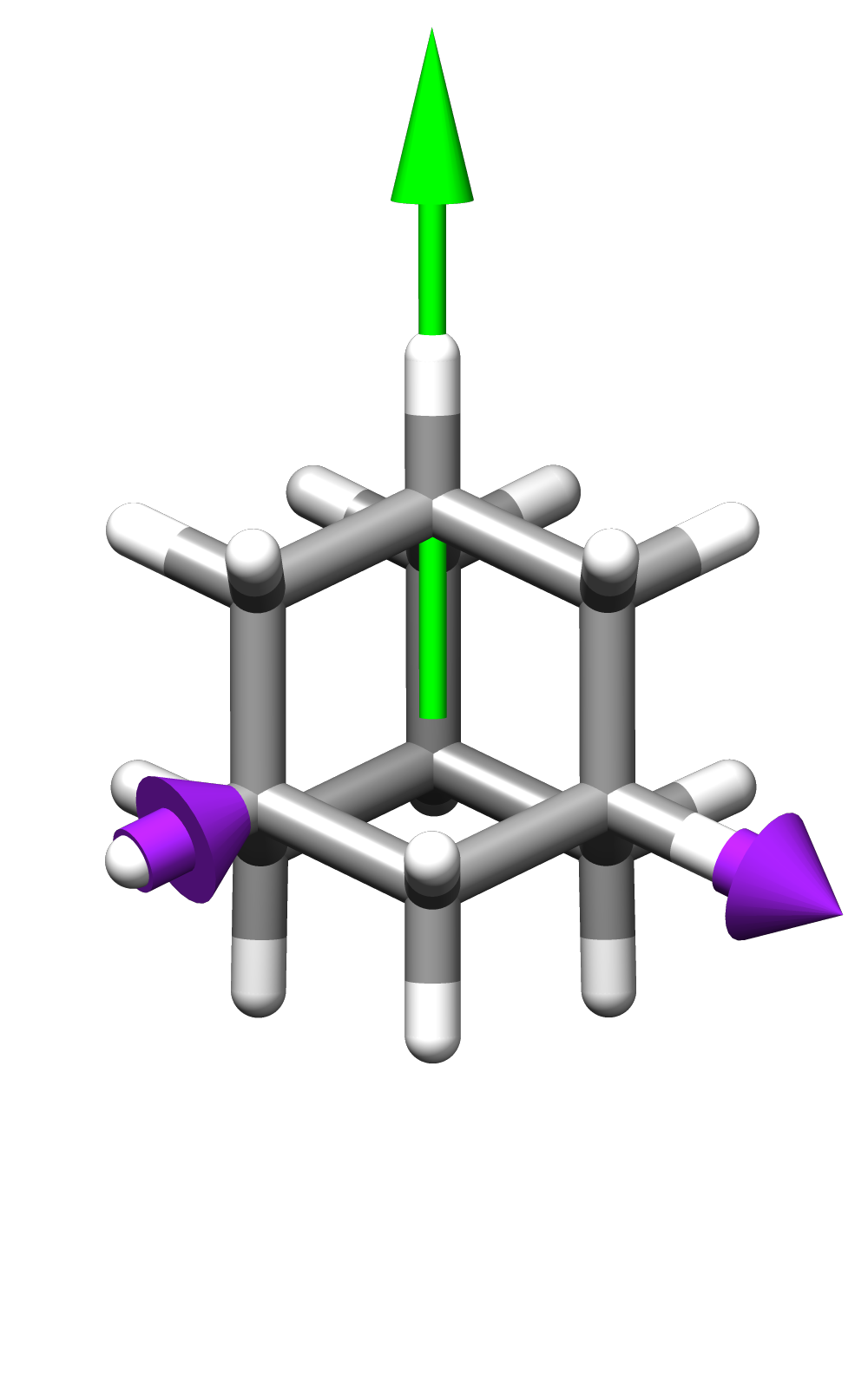}
    \label{fig:ANA_normal_mode_69}
    \end{subfigure}%
    \caption{Normal modes corresponding to the transitions at 3211 cm$^{-1}$ (a) and 3231 cm$^{-1}$ (b) for adamantane at coupling of 0.10 a.u. Purple arrows indicate nuclear displacements, while green arrows represent the polarization orientation of the quantum electromagnetic field.}
\end{figure}

\begin{table}[H]
\centering
\begin{tabular}{crr}
\hline
{Irrep} & {$\tilde{\nu}$ } & {$\mathcal{I}$} \\
\hline
    A$_1$ & 862.1 & 0.44 \\
    E & 865.2 & 0.39 \\
    E & 1213.4 & 3.79 \\
    A$_1$ & 1226.5 & 3.92 \\
    A$_1$ & 1634.4 & 8.42 \\
    E &  1638.6 & 6.81 \\
    E &  3185.6 & 6.70 \\
    A$_1$ & 3188.8 & 36.93 \\
    A$_1$ & 3192.6 & 33.17 \\
    E & 3194.0 & 3.97 \\
    E & 3210.9 & 195.14 \\
    A$_1$ & 3222.8 & 0.0015 \\
    E & 3231.0 & 140.76 \\
    E & 3302.4 & 0.07 \\
    A$_1$ & 3308.2 & 98.94 \\
    A$_1$ & 3314.8 & 108.48 \\
\hline
\end{tabular}
\caption{Vibrational energies (cm$^{-1}$) and IR intensities (km mol$^{-1}$) of adamantane with a coupling of 0.10 a.u.}
\label{tab:ADA_ir_0.10}
\end{table} 
As the coupling with the quantum electromagnetic field increases, nuclear displacements along the polarization axis become more restrained.
This is expected to cause a decrease in the intensity, as observed for the A${_1}$ transition at 3315 cm$^{-1}$. 
This happens when the change in the dipole moment is generated by a nuclear displacement aligned with the quantum electromagnetic field, as shown in Figure \ref{fig:ADA_normal_mode_78} for the aforementioned transition.
In contrast, when nuclear displacements are along other directions, the intensities could be enhanced by the cavity, as observed for the transition to the state to E symmetry at 3211 cm$^{-1}$, whose normal modes are shown in Figure \ref{fig:ANA_normal_mode_69}.

Finally, we note that there is a frequency increase for all transitions due to the contraction of the electronic density induced by the quantum electromagnetic field, as previously observed with formaldehyde and PNA.

\section{Conclusions} \label{sec:conclusions}

In this study, we presented the derivation and implementation of the analytical QED-HF Hessian and explored the influence of the quantum electromagnetic field on molecular vibrational properties. 
We observed significant modifications in vibrational frequencies, intensities, and normal modes induced by strong light-matter coupling. Because our simulations are not carried out in the VSC regime, no polariton formation was observed in the IR spectra. This contrasts with the work by Schnappinger \emph{et al.} \cite{schnappinger2023,schnappinger2023_cbohf}, where they investigated vibrational polaritons. 
Systematic shifts in vibrational frequencies were observed. 
These changes can be attributed to the contraction of electron density, driven by the quantum electromagnetic field. 
In the case of planar systems, the normal modes involving out-of-plane nuclear displacements showed blue-shifted frequencies and reduced intensities. 
As observed in formaldehyde, the combination of such effects causes the wagging and rocking vibrational modes to merge into a single band. 
In PNA, the out-of-plane modes follow the same behavior. In contrast, the wagging vibrational mode of the amino group shows a reduction in frequency and a substantial increase in intensity, indicating the planar geometry as the most stabilized one by the quantum electromagnetic field. 
Strong coupling also impacts molecular symmetry, as observed in adamantane. 
The symmetry reduction from T$_d$ to C$_{3v}$ induced by the quantum electromagnetic field  enables previously forbidden IR transitions to become allowed and leads to the splitting of bands.

Overall, these findings demonstrate that strong coupling significantly influences vibrational frequencies, intensities, and molecular symmetry. 
Furthermore, the analytical evaluation of the QED-HF Hessian offers a promising approach for accurately identifying transition states in cavity-modulated reaction pathways, as chemical reactions are highly responsive to external electromagnetic fields \cite{liebenthal2023assessing}. 
Finally, the analytical evaluation of the Hessian is more efficient and stable than numerical differentiation, paving the way to the calculation of higher-order derivatives, which are required, for instance, for the study of anharmonic effects, which will be explored in a future work.

\section{Data and code availability}

The code used to obtain the findings of this study is available from the corresponding author upon reasonable request. The supporting data are publicly available in ref \citenum{barlini2025_zenodo_qed_hf_hessian}.


\begin{acknowledgement}
    The authors acknowledge funding from the European Research Council (ERC) under the European Union’s Horizon 2020 Research and Innovation Programme (grant agreement no. 101020016).
\end{acknowledgement}

\bibliography{achemso-demo}

\end{document}